\documentstyle[preprint,aps]{revtex}
\begin{document}
%\preprint{KIAS-P99}
\preprint{KNU-H0011}
\def\a{\alpha}
\def\b{\beta}
\def\p{\partial}
\def\m{\mu}
\def\n{\nu}
\def\t{\tau}
\def\s{\sigma}
\def\half{\frac{1}{2}}
\def\hatt{{\hat t}}
\def\hatx{{\hat x}}
\def\hatth{{\hat \theta}}
\def\hatta{{\hat \tau}}
\def\hatrh{{\hat \rho}}
\def\hatva{{\hat \varphi}}
\def\p{\partial}
\def\nn{\nonumber}
\def\cb{{\cal B}}
\def\beq{\begin{eqnarray}}
\def\eeq{\end{eqnarray}}
\def\2pap{2\pi\a^\prime}

\title{Open Superstring and Noncommutative Geometry}
\author{Taejin Lee
\thanks{E-mail: taejin@cc.kangwon.ac.kr}}
\address{{\it
        Department of Physics, Kangwon National University,
        Chuncheon 200-701, Korea}}
\date{\today}
\maketitle
\begin{abstract}
We perform canonical quantization of the open Neveu-Schwarz-Ramond
(NSR) superstrings in the background of a $D$-brane with the
NS B-field. If we choose the mixed boundary condition as a primary
constraint, it generates a set of secondary constraints.
These constraints are easily solved and as a result, the
noncommutative geometry in the bosonic string theory is extended to
the superspace. Solving the constraint conditions we also find that the
Hamiltonian for the superstring is equivalent to a free superstring
Hamiltonian on the target space with the effective open
string metric $G$.
\end{abstract}


\narrowtext

\section{Introduction}
The noncommutative geometry \cite{nc} has been considered for some
time in connection with various physics subjects, which include the
lowest Landau level physics in condensed matter physics, the quantum plane
in mathematical physics, and the geometrical interpretation of
Yang-Mills-Higgs action and so forth.
Recent motivation to study the noncommutative geometry mainly comes from the
string theory. If the matrix model of M-theory \cite{mtheo}
is compactified on tori in the presence of an appropriate
background field, the noncommutative supersymmetric
Yang-Mills theory arises \cite{cds}. It implies that the
$D$-brane world volume theory is noncommutative. This point has
been discussed later by a more direct approach. If we quantize the
open string in the background of a $D$-brane with the NS
$B$-field, the ends points of the string, attached on the
$D$-brane is shown to be noncommutative \cite{direct,cano}. The effective
action for the $D$-brane in the low energy regime is induced by the open
string on it, thus it should be noncommutative.
In their recent paper \cite{sw99} Seiberg and Witten extensively discussed
the various aspects of the noncommutative geometry in the context of the
string theory such as the equivalence between ordinary gauge
fields and the noncommutative gauge fields, Morita equivalence,
and its implications in M-theory.

The noncommutativity in the open string theory can be most easily
seen in the framework of canonical quantization. The approach
based on the canonical quantization was adopted in the earlier works
\cite{cano} on the open string in the $D$-brane background and further
elaborated recently in refs. \cite{tj1,tj2}. In \cite{tj1} we show
that the set of constraints generated by the mixed boundary condition
can be explicitly solved. Solving the constraint conditions we
find the open string is governed by a free string Hamiltonian
defined on the target space with the effective metric $G$.
Resorting to the canonical analysis we also evaluate the Polyakov path
integral for the open string and obtain the noncommutative
Dirac-Born-Infeld action as the low energy effective action for
the $D$-brane, which reduces to the noncommutative Yang-Mills theory
in the zero slope limit. If the constraint conditions are imposed,
the Wilson loop operator with the ordinary gauge field becomes the
Wilson loop operator with a noncommutative gauge field. Hence
the Seiberg-Witten map, which connects the ordinary gauge field to the
noncommutative one, can be understood in the framework of the
canonical quantization \cite{tj2}.

The appearance of the noncommutative geometry in the bosonic
string theory is now well understood. However, the
noncommutative geometry in the framework of the superstring theory needs further study.
Since the $D$-brane is a BPS object, many interesting features
of the $D$-brane are associated
with the supersymmetry. Therefore, it is important to understand
how the noncommutativity interplay with the supersymmetry.
In the present paper we attempt to extend the previous canonical analysis to
the supersymmetric theory. To this end we take the
Neveu-Schwarz-Ramond (NSR) superstring \cite{nsr}
in the background of $D$-brane with NS $B$-field.

\section{The NSR Superstring in the background of $D$-brane}

We begin with the world sheet action for the free NSR superstring
in superspace. In the presence of the $D$-p-brane the action for the
superstring is given as
\beq
S &=& \frac{1}{4\pi\a^\prime}\int_M d^2\xi d^2\theta g_{\m\n}
D_+Z^\m D_-Z^\n \\
Z^\m(\xi, \theta_\pm) &=& X^\m+ \theta_+ \psi^\m_- + \theta_-
\psi^\m_+ +\theta_+\theta_- F^\m \nn\\
D_+ &=& \frac{\p}{\p \theta_-} -i \theta_- \p_+, \quad
D_- = - \frac{\p}{\p \theta_+} +i \theta_+ \p_-  \nn
\eeq
where we choose the world sheet metric as $(-,+)$ and
$F^\m$ is the auxiliary field. $Z^a$, $a=p+1,\dots,9$ of the
transverse directions are subject to the Dirichlet boundary
condition, $\p_\t Z^a=0$.
If the $D$ brane carries the NS $B$-field, the action for the
longitudinal coordinates $Z^i$, $i=0, \dots p$, is replaced
with the following one
\beq
S &=& \frac{1}{4\pi\a^\prime} \int_M d^2\xi d^2\theta
(g+\2pap B)_{ij} D_+Z^i D_-Z^j.
\eeq
Since the action for the transverse coordinates is rather
trivial, we will be concerned with the action for the longitudinal
coordinates only hereafter.
%If we perform the superspace integration and remove the auxiliary
%field, using the on-shell condition,
%\beq
%S &=& \frac{1}{4\pi\a^\prime} \int_M d^2 \xi \Biggl\{ (g+\2pap B)_{ij}
%\p_+ X^i \p_- X^j -i g_{ij} \psi^i_+
%\left(\p_-\delta^j{}_k +\p_- X^j \Gamma^{i+}_{jk}\right)\psi^k \nn\\
%& &-ig_{ij} \psi^i_- \left(\p_+\delta^j{}_k +\p_+ X^j \Gamma^{i-}_{jk}
%\right)\psi^k+ \half R_{ijkl}(\Gamma^-) \psi^i_+ \psi^j_+
%\psi^k_-\psi^l_- \Biggr\}\\
%& & + \frac{i}{4} \int_{\p M} \2pap
%B_{ij} \left(\psi^i_+\psi^j_+ +\psi^i_-\psi^j_-\right) \nn
%\eeq
%where
%\beq
%\Gamma^{i\pm}_{jk} = \Gamma^i_{jk} \pm H^i_{jk},\quad
%H_{ijk} = \half (\p_k B_{ij} +\p_i B_{jk} - \p_j B_{ik})
%\eeq
%and $\Gamma^i_{jk}$ is the Levi-Civita connection.
For the background with constant $g$ and $B$, the action is
greatly simplified to
\beq
S &=& \frac{1}{4\pi\a^\prime}\int_M d^2\xi \left(E_{ij} \p_+X^i \p_-X^j
-i\psi^i_+ E_{ij} \p_-\psi^j_+ -i\psi^i_-
E^T_{ij} \p_+ \psi^j_- \right),\\
E_{ij} &=& (g+\2pap B)_{ij}. \nn
\eeq
The boundary conditions are given as
\begin{mathletters}
\label{bnc:all}
\beq
g_{ij}\p_\s X^i -\2pap B_{ij} \p_\t X^i &=& 0, \label{bnc:a}\\
E_{ij} \psi^j_+ - E^T_{ij} \psi^j_- &=& 0, \label{bnc:b}
\eeq
for $\s = 0$, $\pi$.
\end{mathletters}

For canonical quantization of the bosonic part we refer to
ref. \cite{tj1}: The boundary condition Eq.(\ref{bnc:a}) generates
a set of second class constraints, which can be explicitly solved.
After solving the constraints, the bosonic part of the
Hamiltonian and the string coordinate variables are obtained as
\begin{mathletters}
\label{str:all}
\beq
H_B &=& (2\pi\a^\prime) \half p_i(G^{-1})^{ij}p_j
+(2\pi\a^\prime) \sum_{n=1}\left\{ \half K_{in} (G^{-1})^{ij}
K_{jn} +\frac{1}{(2\pi\a^\prime)^2} \frac{n^2}{2} Y^i_n G_{ij} Y^j_n
\right\}, \label{str:a}\\
X^i(\s) &=& x^i+ \theta^{ij}_{NC}p_j \left(\s - \frac{\pi}{2}\right)+
\sqrt{2} \sum_{n=1} \left(Y^i_n \cos n\s + \frac{1}{n}
\theta^{ij}_{NC} K_{jn} \sin n\s \right) \label{str:b}
\eeq
\end{mathletters}
where $Y^i_n$ and $K^i_n$ satisfy the usual commutation relation
\beq
[Y^i_n, Y^j_m ] =0, \quad
[Y^i_n, K_{jm} ] = i\delta^i{}_j \delta_{nm}, \quad
[K_{in}, K_{jm} ] = 0
\eeq
and
\begin{mathletters}
\label{theta:all}
\beq
\theta^{ij}_{NC} &=& - (2\pi\a^\prime)^2
\left(\frac{1}{g+ 2\pi\a^\prime B} B\frac{1}{g- 2\pi\a^\prime B}
\right)^{ij} \label{theta:a} \\
(G^{-1})^{ij} &=& \left(\frac{1}{g+2\pi\a^\prime B}
g\frac{1}{g-2\pi\a^\prime B} \right)^{ij}. \label{theta:b}
\eeq
\end{mathletters}

Now let us turn to the ferminonic part. The NSR string has
two sectors, depending on the boundary conditions for the
fermionic variables: The Ramond sector with periodic boundary
condition
\beq
\psi^{i+} (\s) = \sum_n \psi^{i+}_n e^{in\s},
\quad \psi^{i-} (\s) = \sum_n \psi^{i-}_n e^{-in\s},
\eeq
and the Neveu-Schwarz sector with anti-periodic boundary condition
\beq
\psi^{i+} (\s) = \sum_n \psi^{i+}_{n+1/2} e^{i(n+1/2)\s},
\quad \psi^{i-} (\s) = \sum_{n} \psi^{i-}_{n+1/2}
e^{-i(n+1/2)\s}
\eeq
where $n \in {\bf Z}$.
Since the constraint Eq.(\ref{bnc:b}) is linear in the fermionic
variables, it is compatible with these boundary conditions.

We discuss the Ramond sector first. The canonical analysis of
the Neveu-Schwarz sector is not much different from that of the Ramond
sector. In the Ramond sector the fermionic part of the action in
the normal modes is written as
\begin{mathletters}
\label{fer:all}
\beq
L_F &=& -i \sum_n \left(\psi^{i+}_n g_{ij}
\dot{\psi}^{j+}_{-n}
+\psi^{i-}_n g_{ij} \dot{\psi}^{j-}_{-n} \right) -H, \label{fer:a}\\
H_F &=& \sum_n n \left(\psi^{i+}_n E_{ij} \psi^{j+}_{-n}
+\psi^{i-}_n E^T_{ij} \psi^{j-}_{-n} \right). \label{fer:b}
\eeq
\end{mathletters}
We can get the Poisson bracket from the this canonical form as
\beq
\{ \psi^{i+}_n, \psi^{j+}_{m}\}_{PB} = i (g^{-1})^{ij}
\delta(n+m),
\quad \{ \psi^{i-}_n, \psi^{j-}_{m}\}_{PB} = i (g^{-1})^{ij}
\delta(n+m).
\eeq
The boundary conditions accordingly read as
\begin{mathletters}
\label{bcf:all}
\beq
\varphi_{0i} &=& E_{ij} \sum_n \psi^{j+}_n - E^T_{ij}
\sum_n \psi^{j-}_n = 0, \label{bcf:a}\\
{\bar \varphi}_{0i} &=&
E_{ij} \sum_n (-1)^n \psi^{j+}_n - E^T_{ij}
\sum_n (-1)^n \psi^{j-}_n = 0. \label{bcf:b}
\eeq
\end{mathletters}

We may choose the boundary condition Eq.(\ref{bcf:a}) as a
primary constraint. Then the Dirac procedure requires
to introduce the following secondary constraint in order to be
consistent
\beq
[ H,\varphi_i ]_{PB} = i \sum_n n \left( E_{ij} \psi^{j+}_n
-E^T_{ij} \psi^{j-}_n \right) = 0.
\eeq
It yields the secondary constraint as
\beq
\varphi_{1i} = \sum_n n \left( E_{ij} \psi^{j+}_n
-E^T_{ij} \psi^{j-}_n \right) = 0.
\eeq
Then the Dirac procedure further requires
\beq
[H, \varphi_{1i} ]_{PB} = 0
\eeq
It leads us to another secondary constraint
\beq
\varphi_{2i} =
\sum_n n^2 \left( E_{ij} \psi^{j+}_n
-E^T_{ij} \psi^{j-}_n \right)=0.
\eeq
We may repeat this procedure until no new secondary constraints
are generated.
By repetition we get
\beq
\varphi_{mi} = \sum_n n^m \left( E_{ij} \psi^{j+}_n
-E^T_{ij} \psi^{j-}_n \right) =0, \,\,\,
m = 1, 2, \dots.
\eeq
Since the obtained constraints are of second class, we should
introduce the Dirac bracket. However, it
may not be convenient to construct the Dirac bracket with
this set of constraints, $\{\varphi_{mi} = 0, m = 1, 2, 3, \dots \}$.
As in the case of the bosonic string theory, the following
observation turns out to be very useful: We can easily disentangle
the set of constraints and find that they are equivalent to
\beq
\left\{E_{ij} \psi^{j+}_n -E^T_{ij} \psi^{j-}_n =0, \quad
n \in {\bf Z}\right\}. \label{cons}
\eeq
(We may also take the boundary condition, Eq.(\ref{bcf:b}), imposed
on the other end of the open superstring. But it generates the
same set of the constraints, Eq.(\ref{cons}); it is redundant.)

The fermionic degrees of freedom are halved by the set of
conditions Eq.(\ref{cons}). They reduce to
\beq
\left\{\psi^{i+}_n - \psi^{i-}_n =0, \quad n \in {\bf Z}\right\}
\eeq
when the NS $B$-field is absent.
In the case of the free superstring theory we get rid
of $\psi^{i-}_n$ in favor of $\psi^{i+}_n$ and
choose $\{\psi^{i+}_n\}$ as a
proper basis for the fermionic degrees of freedom. Suppose that we
choose $\{\psi^{i+}_n\}$ as the basis for the
fermionic degrees of freedom in the present case.
If we make use of the constraints,
$\psi^{i-}_n = ((E^T)^{-1} E)^i{}_j \psi^{j+}_n$,
we find that the fermionic part of the Lagrangian is written as
\beq
L_F &=& -i \sum_n \left(\psi^{i+}_n g_{ij} {\dot \psi}^{j+}_{-n}\right)
- H_F, \\
H_F &=& \sum_n n \left(\psi^{i+}_n g_{ij} \psi^{j+}_{-n}\right)\nn
\eeq
where we make use of
\beq
E^T E^{-1} g (E^{-1})^T E =E^T G^{-1} E = g.
\eeq
Here $\psi^{i+}$ is scaled as $\psi^{i+}\rightarrow
\psi^{i+}/\sqrt{2}$.
At first appearance the Lagrangian looks same as that of the free
superstring and the NS $B$-field does not affect the fermionic
part. However, this conclusion is misleading. If the constraint
conditions are imposed the bosonic part respects the effective
metric $G$ instead of $g$. If the fermionic part still respects
the metric $g$ after imposing the constraints, the super-Virasoro
algebra would not be consistent. We will discuss this point
in the next section in some detail.

\section{Supersymmetry and Noncommutative Geometry}

Let us recall the bosonic part of the Hamiltonian
for the open string in the background of NS $B$-field.
If the background NS $B$-field is constant, the Hamiltonian
is given as
\begin{mathletters}
\label{hamb:all}
\beq
H^B &=& \half\left(L^B_0 + {\bar L}^B_0\right), \label{hamb:a}\\
L^B_0 &=& (2\pi\a^\prime) p_{Li} (g^{-1})^{ij} p_{Lj}+
(2\pi\a^\prime) \sum_{n=1} n A_{in} (g^{-1})^{ij} A^\dagger_{jn}, \label{hamb:b}\\
{\bar L}^B_0 & =& (2\pi\a^\prime) \half p_{Ri} (g^{-1})^{ij} p_{Rj}+
(2\pi\a^\prime) \sum_{n=1} n {\bar A}_{in}
(g^{-1})^{ij} {\bar A}^\dagger_{jn}, \label{hamb:c}
\eeq
\end{mathletters}
where
\begin{mathletters}
\label{p:all}
\beq
p_{Li} &=& \frac{1}{\sqrt{2}}\left(p_i -
\frac{1}{\2pap} E^T_{ij}a^j\right), \label{p:a}\\
p_{Ri} &=& \frac{1}{\sqrt{2}}\left(p_i +
\frac{1}{\2pap} E_{ij}a^j\right). \label{p:b}
\eeq
\end{mathletters}
The left and right movers are defined as
\begin{mathletters}
\label{mover:all}
\beq
A^\dagger_{in} &=& \frac{1}{\sqrt{2n}} \left(P_{i-n}+
\frac{in}{\2pap} E^T_{ij}X^j_n\right), \label{mover:a}\\
A_{in} &=& \frac{1}{\sqrt{2n}} \left(P_{in}-\frac{in}{\2pap}
E^T_{ij}X^j_{-n}\right), \label{mover:b}\\
{\bar A}^\dagger_{in} &=&
\frac{1}{\sqrt{2n}} \left(P_{in}+\frac{in}{\2pap}
E_{ij}X^j_{-n}\right), \label{mover:c}\\
{\bar A}_{in} &=& \frac{1}{\sqrt{2n}} \left(P_{i-n}-
\frac{in}{\2pap} E_{ij}X^j_n\right), \label{mover:d}
\eeq
\end{mathletters}
and their commutation relations are
\beq
[A_{in}, A^\dagger_{jm}] =
(2\pi\a^\prime)^{-1} g_{ij} \delta_{nm},\quad
[{\bar A}_{in}, {\bar A}^\dagger_{jm}] =
(2\pi\a^\prime)^{-1} g_{ij} \delta_{nm}.
\eeq
Using the constraint conditions
\beq
a^i = \theta^{ij}p_j, \quad {\bar Y}^i_n = \frac{1}{\sqrt{2}}(X^i_n-X^i_{-n})
= \frac{1}{n} \theta^{ij}K_{jn}, \quad
{\bar K}_{in} = \frac{1}{\sqrt{2}}(P^i_n-P^i_{-n}) = 0,
\eeq
we may remove $a^i$,
${\bar Y}^i_n$, ${\bar K}_{in}$ in favor of $(x^i, p_i)$,
$(Y^i_n, K_{in})$, which are canonical variables for
the open string. As a result we find that the bosonic part of
the Hamiltonian is just the same as the free Hamiltonian
for the open string on the target space with metric
$G$
\begin{mathletters}
\label{ham2:all}
\beq
H^B &=& (2\pi\a^\prime) \half \left(p^2 + \sum_{n=1} n
A_{in}(E^\prime) \left(G^{-1}\right)^{ij}
A^\dagger_{jn}(E^\prime)\right), \label{ham2:a}\\
A_{in}(E^\prime)&=& \frac{1}{\sqrt{2n}} \left(K_{in}-\frac{in}{\2pap}
G_{ij}Y^j_{n}\right), \label{ham2:b}\\
A^\dagger_{in}(E^\prime)&=& \frac{1}{\sqrt{2n}} \left(K_{in}+
\frac{in}{\2pap} G{ij}Y^j_n\right), \label{ham2:c}
\eeq
\end{mathletters}
where the left movers and the right movers satisfy
\beq
[ A_{in}(E^\prime), A^\dagger_{jm}(E^\prime) ] =
(\2pap)^{-1} G_{ij} \delta(n+m).
\eeq
It is interesting to note \cite{tj3} that the left and
right movers $A^\dagger(E)$, ${\bar A}^\dagger(E)$ are
related to the left and right movers, $A^\dagger(E^\prime)$,
${\bar A}^\dagger(E^\prime)$ by a T-dual transformation
\cite{tdual}
\beq
T =\left( \begin{array}{cc} I & 0 \\
(\2pap)^{-1} \theta & I \end{array} \right).
\eeq
We may have geometric interpretation of this T-dual
transformation as discussed in ref. \cite{ima}.

Now let us turn to the fermionic constraint $F_0$,
which forms the super-algebra with the Hamiltonian $H$,
\beq
\{F_0, F_0\} + \{{\bar F}_0, {\bar F}_0\}=
2 \left(L_0+ L^F_0\right)+ 2 \left({\bar L}_0+
{\bar L}^F_0\right)
= 2 \left(H^B+ H^F\right).
\eeq
In the presence of the NS $B$-field the fermionic constraint
$F_0$ is given as
\begin{mathletters}
\label{ferc:all}
\beq
(\2pap)^{-\half}F_0 &=& p_{Li} \psi^{i+}_0 +
\sum_{n=1}  \sqrt{n} \left(A_{in} \psi^{i+}_{-n} +
A^\dagger_{in} \psi^{i+}_n \right), \label{ferc:a}\\
(\2pap)^{-\half}{\bar F}_0 &=& p_{Ri} \psi^{i-}_0 +
\sum_{n=1}  \sqrt{n} \left( {\bar A}_{in} \psi^{i-}_{-n} +
{\bar A}^\dagger_{in} \psi^{i-}_n \right). \label{ferc:b}
\eeq
\end{mathletters}
From the canonical analysis of the bosonic part we expect
that the fermionic constraint is rewritten as
\beq
(\2pap)^{-\half}\left( F_0+{\bar F}_0\right) =
p_i \hat{\psi}^{i}+ \sum_{n=1} \sqrt{n}
\left(A^\dagger_{in}(E^\prime) \hat{\psi}^{i}_n +
A_{in}(E^\prime) \hat{\psi}^{i}_{-n}\right)
\eeq
where the fermion operator $\hat{\psi}^{i}_n$ satisfies
\beq
\{ \hat{\psi}^{i}_n, {\hat \psi}^{j}_{m}\}_{PB} =
(G^{-1})^{ij} \delta(n+m).
\eeq
It follows then that the fermion operator $\hat{\psi}^{i}_n$
may be defined as
\beq
\hat{\psi}^i_n = \frac{1}{\sqrt{2}} \left(\psi^{i+}_n
+ \psi^{i-}_n \right) = \sqrt{2} (G^{-1}E)^i{}_j \psi^{j+} =
\sqrt{2} (G^{-1}E^T)^i{}_j \psi^{j-}.
\eeq
Rewriting the fermionic part of the Lagrangian Eq.(\ref{fer:all}) as
\begin{mathletters}
\label{lag:all}
\beq
L_F &=& -i \sum_n \left({\hat \psi}^{i}_n G_{ij} \p_\t
{\hat \psi}^{j}_{-n}\right) - H_F, \label{lag:a}\\
H_F &=& \sum_n n \left({\hat \psi}^{i}_n G_{ij}
{\hat \psi}^{j}_{-n}\right), \label{lag:b}
\eeq
\end{mathletters}
we confirm that the fermionic part also respects the
effective open string metric $G$.
It is consistent with the commutation relation among
$\{{\hat \psi}^{i}_n\}$ and the supersymmetry.

Being equipped with the canonical analysis of the fermionic
part, we discuss the noncommutativity in the superspace
\beq
Z^i (\s) = X^i(\s)+ \theta_+ \sum_n \psi^{i-}_n e^{-i n\s}+
\theta_-\sum_n \psi^{i+}_n e^{i n\s} + \theta_+\theta_- \sum_n
F^i e^{in\s}.
\eeq
We may rewrite the fermionic part of $Z^i$ in terms of
${\hat \psi}^{i}_n$ as
\beq
Z^i_F &=& \frac{1}{\sqrt{2}} \sum_n \left\{
\left(\theta \cos n\s - i {\bar \theta} \sin n\s \right)
{\hat \psi}^{i}_n + i
\frac{\left(\theta_{NC} G\right)^i{}_j}{\2pap}
\left(\theta \sin n\s +i {\bar \theta} \cos n\s \right)
{\hat \psi}^{j}_n\right\}
\eeq
where
\beq
\theta = \frac{1}{\sqrt{2}} (\theta_+ +\theta_-), \quad
{\bar \theta} =  \frac{1}{\sqrt{2}} (\theta_+ -\theta_-), \quad
{\bar \theta} \theta = \theta_+ \theta_- . \nn
\eeq
Note that the end points of the string are no longer
holomorphic in $\theta$
\begin{mathletters}
\label{zi:all}
\beq
Z^i_F(0) &=& \frac{1}{\sqrt{2}} \sum_n \left(
\theta {\hat \psi}^{i}_n - {\bar \theta}
\frac{\left(\theta_{NC} G\right)^i{}_j}{\2pap}
{\hat \psi}^{j}_n\right), \label{zi:a}\\
Z^i_F(\pi) &=& \frac{1}{\sqrt{2}} \sum_n (-1)^n \left(
\theta {\hat \psi}^{i}_n - {\bar \theta}
\frac{\left(\theta_{NC} G\right)^i{}_j}{\2pap}
{\hat \psi}^{j}_n\right).\label{zi:b}
\eeq
\end{mathletters}
It may have some consequences in construction of the vertex
operators. The noncommutativity can be easily seen
if we evaluate the commutator between $Z^i_F$
\beq
[Z^i(\s), Z^j(\s^\prime)] = [X^i(\s), X^j(\s^\prime)] +
[Z^i_F(\s), Z^j_F(\s^\prime)].
\eeq
The noncommutativity in the bosonic sector is discussed in
details in the previous works \cite{cano,tj1}.
The commutation relation between $Z^i_F(\s)$ is found to be
\beq
[Z^i_F(\s), Z^j_F(\s^\prime)]
&=& \left\{ \begin{array}{r@{\quad:\quad}l}
\frac{i}{2\pi\alpha^\prime} (\theta {\bar \theta}) \theta^{ij}_{NC}
 & \s=\s^\prime=0, \\
-\frac{i}{2\pi\alpha^\prime} ({\bar \theta}\theta) \theta^{ij}_{NC}
 & \s=\s^\prime=\pi \\
0 & {\rm otherwise} \end{array} \right.
\eeq
where we use $\zeta(0)= \sum_n 1= -1/2$.

The canonical analysis of the Neveu-Schwarz Sector is obtained
as we replace the integer modes of the fermion variables
by the half-integer modes, $\psi_n \rightarrow \psi_{n+1/2}$.

\section{Concluding Remarks}

A few remarks are in order to conclude the paper.
The first one is a brief summary of the present work.
We perform the canonical quantization of the
NSR open superstring
attached on the $D$-brane with a NS $B$-field.
The open superstring has fermionic boundary conditions
to be imposed on the ends of the string in addition to the
bosonic ones. Taking the fermionic boundary conditions
as primary ones, we obtain a set of infinite secondary
constraints, which turn out easy to solve.
Choosing an appropriate basis for the fermion variables,
we find that the fermion variables also respect the
effective metric $G$ and the fermionic part of the Hamiltonian
is just same as that of the free open string Hamiltonian
defined on the target space with the effective metric $G$.
Thus, the canonical analysis of the bosonic part is
extended to the fermionic part explicitly. The interesting
point we should note is that the superspace coordinate
$Z^i$ is no longer holomorphic in $\theta$ at the end points
of the string. We expect that this has some important
consequences when we construct the vertex operators
representing physical states.

The present canonical analysis suggests a number of interesting
directions to explore along the line of this work.
We may construct the vertex operator for emission of
a scalar to study the recent issues associated with
the noncommutative field theories \cite{ncft} in the context of
superstring theory. The supersymmetric Dirac-Born-Infeld
action \cite{sudbi} may be derived if we construct
the vertex operator for emission of a massless vector and
evaluate the Polyakov string path integral over a disk
on the D-brane word sheet. It would also bring us to the
Seiberg-Witten map in the context of the
supersymmetric noncommutative field theory.
The T-duality also deserves further study and the Morita
equivalence may be extended to the supersymmetric theory.
After all these related subjects may be discussed in a
single framework of the second quantized open superstring
theory, if properly constructed. The canonical analysis
presented here would be certainly useful to develop the
second quantized open superstring theory on the noncommutative
geometry.

\section*{Acknowledgement}
This work was supported in part by KOSEF (995-0200-005-2).
Part of the work was done during the author's visit to KIAS.

\end{document}